\pdfoutput=1
\documentclass[12pt]{article}

\usepackage{graphicx}
\usepackage{graphics}
\usepackage{verbatim}   
\usepackage{color}      
\usepackage{subfigure}  
\usepackage{amssymb}    
\usepackage{amsmath}    
\usepackage{amsthm}     
\usepackage[dvips]{epsfig}

\usepackage{hyperref}

\usepackage{float} 
\restylefloat{table}

\usepackage{booktabs} 

\usepackage{enumitem} 

\usepackage{rotfloat} 

\usepackage{placeins} 

\newcommand{\blind}{0}

\begin{document}
	
	\def\spacingset#1{\renewcommand{\baselinestretch}%
		{#1}\small\normalsize} \spacingset{1}

\if0\blind
{
	\title{\bf Incorporating Open Data into Introductory Courses in Statistics}
	\author{Roberto Rivera\thanks{College of Business, University of Puerto Rico, Mayaguez} \and  Mario Marazzi\thanks{Puerto Rico Institute of Statistics} \and  Pedro A. Torres-Saavedra\thanks{Department of Mathematical Sciences,  University of Puerto Rico, Mayaguez}}
	\maketitle
} \fi

\if1\blind
{
	\bigskip
	\bigskip
	\bigskip
	\begin{center}
		{\LARGE\bf Incorporating Open Data into Introductory Courses in Statistics}
	\end{center}
	\medskip
} \fi




\pagestyle{plain} 


\bigskip
\begin{abstract}
	The 2016 Guidelines for Assessment and Instruction in Statistics Education (GAISE) College Report emphasized six recommendations to teach introductory courses in statistics. Among them: use of real data with context and purpose. Many educators have created databases consisting of multiple data sets for use in class; sometimes making hundreds of data sets available. Yet `the context and purpose' component of the data may remain elusive if just a generic database is made available.
	
	We describe the use of open data in introductory courses. Countries and cities continue to share data through open data portals. Hence, educators can find regional data that engages their students more effectively. We present excerpts from case studies that show the application of statistical methods to data on: crime, housing, rainfall, tourist travel, and others. Data wrangling and discussion of results are recognized as important case study components. Thus the open data based case studies attend most GAISE College Report recommendations. Reproducible \textsf{R} code is made available for each case study. Example uses of open data in more advanced courses in statistics are also described. 
\end{abstract}

\noindent%
\vfill

\newpage

\section{Background}
In 2016, the eleven-year-old GAISE College Report was revised. There were two reasons for the revision, the increase in available data and the emergence of data science as a discipline (\cite{gaise16}). The GAISE College report recommends the use of real data with context and purpose. The report even mentions the New York City open data portal as a reference. However, almost all data sets used as example or referred to in the report can not be considered regional for most course students. Furthermore, most data sets referred to in the report contain fewer than a few thousand observations. 

Two important obstacles to effective statistical education for non-statistics majors that receive little attention from statisticians are statistics anxiety and attitude toward statistics. Statistics anxiety has been defined as the feelings of anxiety encountered when taking a
statistics course or doing statistical analysis (\cite{cruiseetal85}). Attitude toward statistics is an individual’s disposition to respond either
favorably or unfavorably to statistics or statistics learning (\cite{chewanddillon15}). It has been found that a negative attitude towards statistics results in statistics anxiety (\cite{chewanddillon14}). Several studies have found that there is a negative association between statistics anxiety and achievement in statistics courses (\cite{chewanddillon14}), although some studies have suggested that some statistics anxiety (but not too much) may be beneficial to students (see \cite{onwuegbuzieandwilson03,keeleyetal08}). Statistics anxiety also affects non-statistics major graduate students (\cite{williams10}). The influence of statistics anxiety on student achievement has led to several recommendations, among them using real data (\cite{neumannetal13}), and the reduction of mathematical emphasis (\cite{chewanddillon14}).



The curriculum must prepare students to engage in the entire data analysis process including data wrangling (\cite{hortonandhardin15}). With the emergence of big data, data science and data analytics, more emphasis in large data is needed in courses. \cite{ridgway16} suggests devoting course time to open data. \cite{manyikaetal13} define open data as information available to everyone; at zero cost; without limits of reuse and redistribution; and are machine readable (in formats that can be easily retrieved and processed by computers). However, this definition leaves out several open data traits that are important when teaching statistics: 
\begin{itemize}
	\item Open data is current and localized, allowing to incorporate current hot topics into the course.
	\item Filtering, aggregating, cleaning and other pre-processing could be needed; presenting an opportunity to introduce students to data management and data wrangling.
	\item Open data can be very large, and even massive.
\end{itemize}
In contrast, other data available for courses are rarely current; are always neat and ready for use; and rarely exceed a few hundred observations (\cite{baumer15,grimshaw15}).

In this paper, we present a series of uses of open data in class. We argue that in terms of student exposure, open data sets are an excellent resource. Challenges of bringing open data into the classroom are also discussed.

\section{Case Studies}
\label{sec:cs1pol}
A set of case studies revolving around open data are covered in this section. Usually, students are presented with the problem addressed in the case study and asked what they think the solution or answer should be before seeing the results of the statistical procedure.  Later students are introduced to the data. Even if the data is downloaded and made accessible to the students, we encourage instructors to briefly show the students the data from the website so that they can appreciate the authenticity of the observations. Any data wrangling is either completed together with the class or pointed out as an already performed task. One final remark is that for presentation and reproducibility purposes, all statistical procedures in this paper were performed through \textsf{R} (\cite{rcran16}). Codes are available as supplementary material. Statistics majors can use the codes to replicate the results or modify them for a different purpose. However, considering the influence of statistical anxiety on the performance of non-statistics majors, we recommend instructors rely on a more user friendly software than \textsf{R} for any coursework. We have successfully required students to use Minitab, and visualization tools (e.g. dashboards)
accessible through many open data portals.

\subsection{Obtaining a metric of violence in Puerto Rico}\label{sec:prm}
\begin{figure}[H]
	\begin{center}

		\includegraphics[width=3.8in]{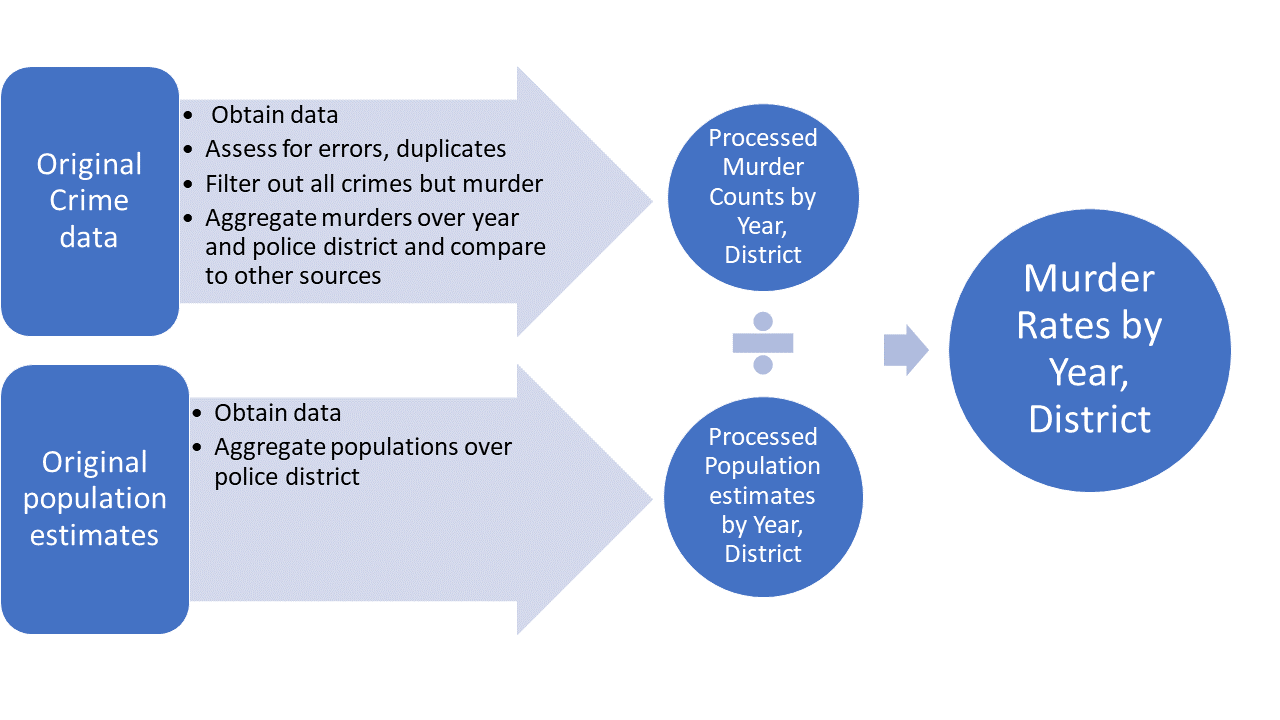}
	\end{center}
	\caption{Steps required to calculate murder rate data for Puerto Rico by year and police district.} 
	\label{fig:murderratesteps}
\end{figure}

The learning objectives of this case study are:
\begin{itemize}
	\item Understand how data wrangling can be applied.
	\item Interpret computer generated descriptive statistics.
	\item Evaluate the limitations of the case study.  
\end{itemize}
How can we compare violence in one place to violence in another? At first we simply
state we will use murder data; a natural choice, to compare violence between multiple locations. Students tend to quickly realize that such a comparison should not be solely based on the number of murders; the demographics of the places to be compared play a role. In this case study, Puerto Rico murder rates per 100,000 people are computed from open data and compared among several regions. As a first look at data science, students are told that the murder rates were obtained by processing two data sets\footnote{Sources: \url{https://data.pr.gov/api/views/pzaz-tkx9/rows.csv}, \url{https://www.indicadores.pr/dataset/estimados-anuales-poblacionales}} from two open data portals (see Figure \ref{fig:murderratesteps}). The crime data used includes dates, times and location of 9 different
types of crimes. As of this writing, the raw crime data set had over 250,000 rows\footnote{\url{https://data.pr.gov/Seguridad-P\%C3\%BAblica/Mapa-del-Crimen-Crime-Map/bkiv-k4gu}}. Only data from 2012 to 2015 was considered, since 2016 data was incomplete. Furthermore, even if data would have been available until the end of 2016, many current cases were still under investigation. Vintage 2016 U.S. Census annual population estimates for the period of interest were also retrieved \footnote{\url{ https://indicadores.estadisticas.pr/dataset/2eabdd0d-625b-4936-af25-5d0258e86ae4/resource/8a2b3a66-7248-4c21-9631-545df1ffc2d9/download/estimados_anuales_poblacionales_por_municipio_y_puerto_rico.csv}}. The quality of the crime
data was assessed via checking for duplicated entries and comparison of annual
number of murders found in other sources. 

Initially a comparison of yearly murder rates in four police districts (San Juan, Fajardo, Ponce and Mayaguez) in Puerto Rico was performed. It was found that San Juan had the highest murder rate among the four police districts, with results of 49.6, 42.2, 43.1, and 33.2 murders per 100,000 people for 2012, 2013 2014, and 2015 respectively. As in the other three regions, the murder rate in San Juan had been decreasing. Without prior knowledge about murder rates, it is still hard to grasp the meaning of the San Juan murder rates. Thus, San Juan murder rates are compared next to cities of at least 250,000 people in the United States. To get this stage of the case study going, students are asked which U.S. cities have the highest murder rates. New York City, Los Angeles and Chicago are frequently mentioned. Next, the audience is presented with Table \ref{tab:murderratestop5}, which lists the cities with the five highest 2015 murder according to the 2015 FBI violent crime report\footnote{Source: \url{https://ucr.fbi.gov/crime-in-the-u.s/2015/crime-in-the-u.s.-2015/tables/table-6}}, San Juan would place fifth highest among all large cities. 

\begin{table}[H]
	\centering
	\begin{tabular}{ccccc}
		\hline
		\textbf{St. Louis} & \textbf{Baltimore} & \textbf{Detroit}      & \textbf{New Orleans}      & \textbf{San Juan}   \\ 
		59.2& 55.37 &43.82               & 41.68               & 33.2   \\ \hline
	\end{tabular}
	\caption{Top five 2015 murder rates (per 100,000 people) in U.S. cities and San Juan, Puerto Rico.}\vspace{2mm}
	\label{tab:murderratestop5}
\end{table}

Given the status of Puerto Rico as a commonwealth, one can compare the overall island murder rate with U.S. states and with other countries as well. Table \ref{tab:PRUS2} summarizes the murder rates for Puerto Rico, New York, Florida and California. The statistics for the island are always over 3 times higher than these benchmarks. Moreover, the 2014 murder rate in Puerto Rico was almost twice as high as that of Louisiana, which was the state with the highest murder rate in the U.S. back then.
\begin{table}[H]
	\centering
	\begin{tabular}{lcccc}
		\hline
		\textbf{Year} & \textbf{Puerto Rico} & \textbf{Nueva York} & \textbf{Florida} & \textbf{California} \\ \hline
		\textbf{2012} & 26.9  & 3.5 & 5.2  & 5.0   \\ 
		\textbf{2013} & 25.3  & 3.3 & 5.0 & 4.6  \\ 
		\textbf{2014} & 19.3  & 3.1  & 5.8  & 4.4  \\ 
		\textbf{2015} & 16.9  & --  & --  & --     \\ \hline
	\end{tabular}
\caption{Murder rate per 100,000 residents in some of U.S. states and Puerto Rico:}\vspace{2mm}
\label{tab:PRUS2}
\end{table} 

The United States' overall 2015 murder rate was about 4.9 murders per 100,000 people, up 10\% from the previous year, yet still far below the 16.9 murders per 100,000 people in Puerto Rico. Now, it has to be highlighted the United States is a large country and there is considerable variability in murder rates within it. Also, gun violence is considered an important public health challenge in the country (\cite{kalesanandgalea17}). To further put things in perspective, many European countries have murder rates below 1 murder per 100,000 people, while Honduras had 84.6 murders per 100,000 people in 2014. In summary, the good news is that the computed murder rates indicate a decrease in violence in Puerto Rico; but much work is needed. The case study is finalized with a discussion of the results, including the limitations of the statistical procedure. Among the limitations: no crime expert was involved in the study, murder is not the only measure of violence (e.g. violent crime data can also be used, \cite{short18}), and variables such as population density and economic hardship play a role in violence, but were not included in the presentation.

\subsection{Do police searches during vehicle stops in San Diego depend on driver's race?}\label{sec:sdpol}
The learning objectives of this case study are:
\begin{itemize}
	\item Recognize that data wrangling is often needed.
	\item Interpret bar plots.
	\item Evaluate the limitations of the case study.  
\end{itemize} 
Students are presented with a case study aiming to visually assess the question that is the title of this section. Students are allowed to argue why they think police searches depend or do not depend on race before using the data. The original San Diego vehicle stop data is shown to the students\footnote{\url{https://data.sandiego.gov/datasets/police-vehicle-stops/}}. Whether gender has any impact in the results is also assessed. Observations with missing values in race, search, or gender were removed from the data set. 

The statistical concept to be taught is data visualization, specifically bar plots, summarizing multiple categorical variables. However, this case study can very well be applied in the context of probability, or inference on two categorical variables. The original data consists of over 100,000 incidents of vehicle stops. After exploring the data, one of the noticeable things is that the race variable is rather specific (Korean, Japanese, Indian, etc.). To simplify our task, race is reclassified into 5 categories: Blacks, Hispanics, Whites, Asians and Others.

\begin{figure}[H]
	\begin{center}

		\includegraphics[width=3.8in]{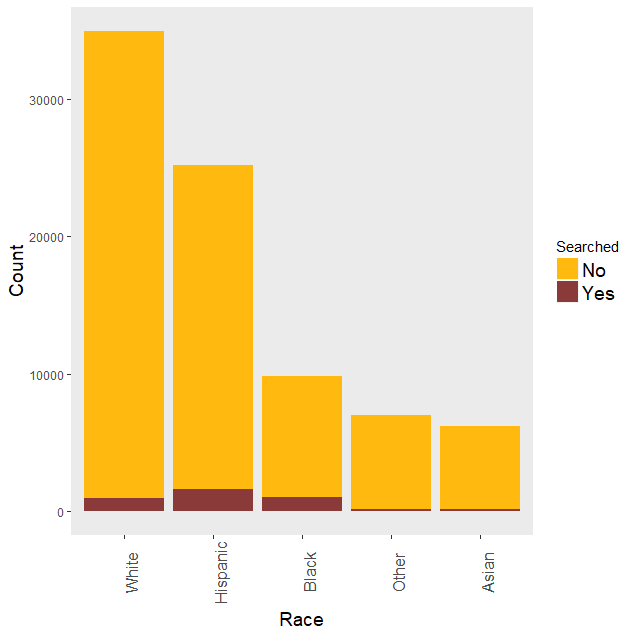}
	\end{center}
	\caption{Number of vehicle stops and searches by race in San Diego in 2016.} \label{fig:tstopsSDoverall}
\end{figure}

Figure \ref{fig:tstopsSDoverall} summarizes the 2016 San Diego vehicle stops by race and whether the
driver was searched. Overall, White drivers were stopped the most, but it is explained to students that this does not mean that White drivers are more likely to be stopped by police, than drivers from other races (according to race distribution, we should expect that the majority of drivers stopped are White or Hispanic). Assessing likeliness of being stopped by race could benefit from using other data such as time of day of vehicle stops (during the day it is easier to identify the race of a driver). Drivers who are Black or Hispanic appear to be searched more often than drivers from other races. Specifically, for Black drivers the red bar covers more of the entire race bar than for White drivers. The same can be said for Hispanic drivers. 

Figure \ref{fig:tstopsSD} summarizes the 2016 San Diego vehicle stops by race, gender and whether the
driver was searched. The class is asked to interpret the stacked bar chart, at first comparing genders and then just focusing on men. In summary:
\begin{figure}[H]
	\begin{center}

		\includegraphics[width=3.8in]{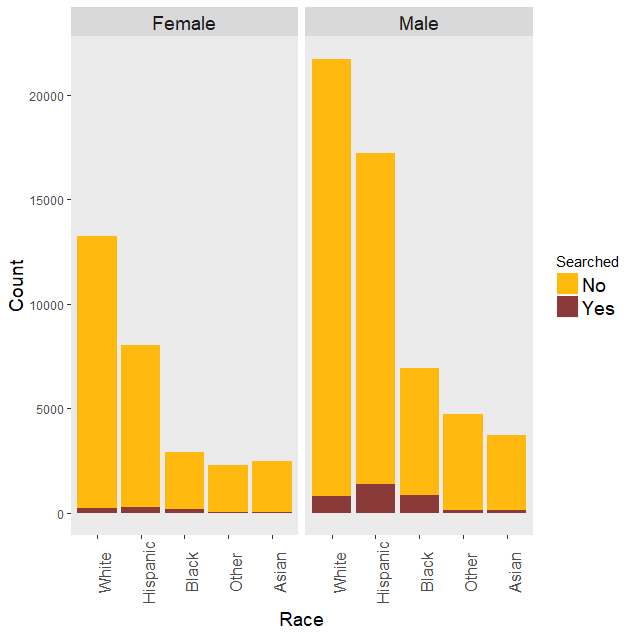}
	\end{center}
	\caption{Number of vehicle stops and searches by race and gender in San Diego in 2016.} \label{fig:tstopsSD}
\end{figure} 
\begin{itemize}
	\item Figure \ref{fig:tstopsSDoverall} indicates that searches occur more frequently when the driver is Black or Hispanic.
	\item Overall, men are generally stopped more frequently than women, with White
	men being stopped the most (considering that women may be more likely than men to experience statistics anxiety (\cite{chewanddillon14, balougluetal11}), and humor has been proposed as a way to reduce statistics anxiety (\cite{chewanddillon14, williams10}), this is a perfect opportunity for humor; stating that the claim that women are bad drivers is not supported by the data). 
	\item Drivers from `Asian' and `Other' races had the least stops and least searches.
\end{itemize}

To have a
better idea of what is suggested by these bar charts, numbers are needed. The percentage of stopped drivers who were subsequently
searched, according to whether they were Black, Hispanic or White were 12.4\%, 7.84\%, 3.57\% respectively. 

Limitations of the completed statistical procedure must be pointed out. The assessment does not answer why there is such a discrepancy in searches of drivers by race. For
one, the summary does not consider the cause for the vehicle stop, or location in San Diego. Also, the
data does not include the race of the officer which may (or may not) be associated to the chance
that a vehicle stop involves a search. Furthermore, statistical inference would be needed to draw
conclusions from the current data since the discrepancy by race could be due to random chance. The presentation is followed by a discussion.

\subsection{Other Open Data Based Ideas}

\begin{figure}[H]
	\begin{center}		
		\includegraphics[width=3.8in]{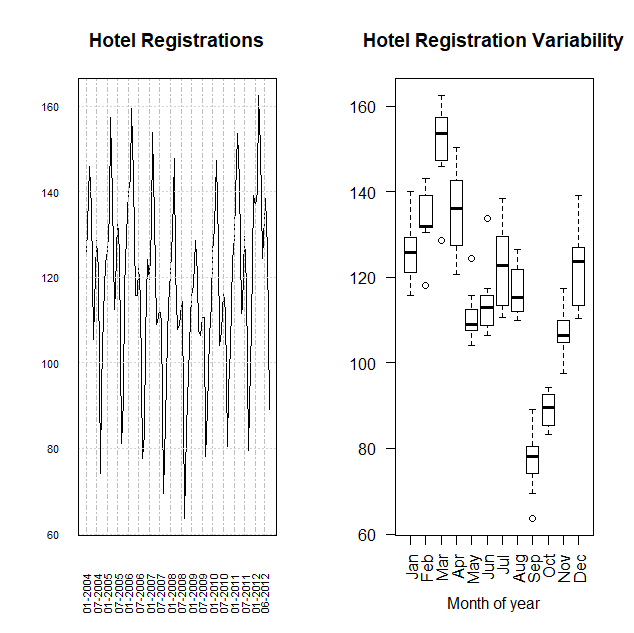}
	\end{center}
	\caption{Left panel displays the time series plot of monthly nonresident hotel registrations in Puerto Rico (in thousands). Right panel shows the separate distributions of monthly
		nonresident registrations (in thousands) for each calendar month.} \label{fig:tsboxplot}
\end{figure}

\begin{figure}[H]
	\begin{center}		
		\includegraphics[width=3.8in]{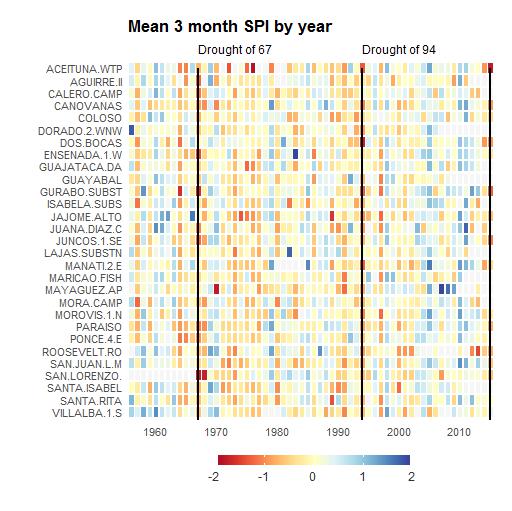} 
	\end{center}
	\caption{Annual average of Standardized Precipitation Index (SPI), \cite{mckeeetal93}, for accumulation of rain every 3 months across many weather stations in Puerto Rico.  Each line represents a station and each column indicates a year.  Period covered is 1956 thru 2015. Negative SPI indicates below average rain at the weather station while positive SPI indicates above average rain. Vertical lines for droughts of 1967, 1994, and 2015 serve as benchmarks.} \label{fig:meanyearly3monthSPI}
\end{figure}

Many other case studies can be created using open data sets. An introduction to time series analysis could be presented using monthly hotel registration data (Figure \ref{fig:tsboxplot}). Data visualization has received a lot of attention lately (\cite{hullmanetal15,nolanandperrett16}). Hans Rosling (\cite{oneill18}) showed how effective data visualization can be in engaging the audience. More frequently, traditional graphics are being extended by adding additional dimensions: more variables (Figure \ref{fig:Baltimoremotionchartnewvar}. See supplementary material for the interactive version). This has made displaying complicated information more aesthetically appealing (Figure \ref{fig:meanyearly3monthSPI}). 

\begin{figure}[H]
	\begin{center}

		\includegraphics[width=3.8in]{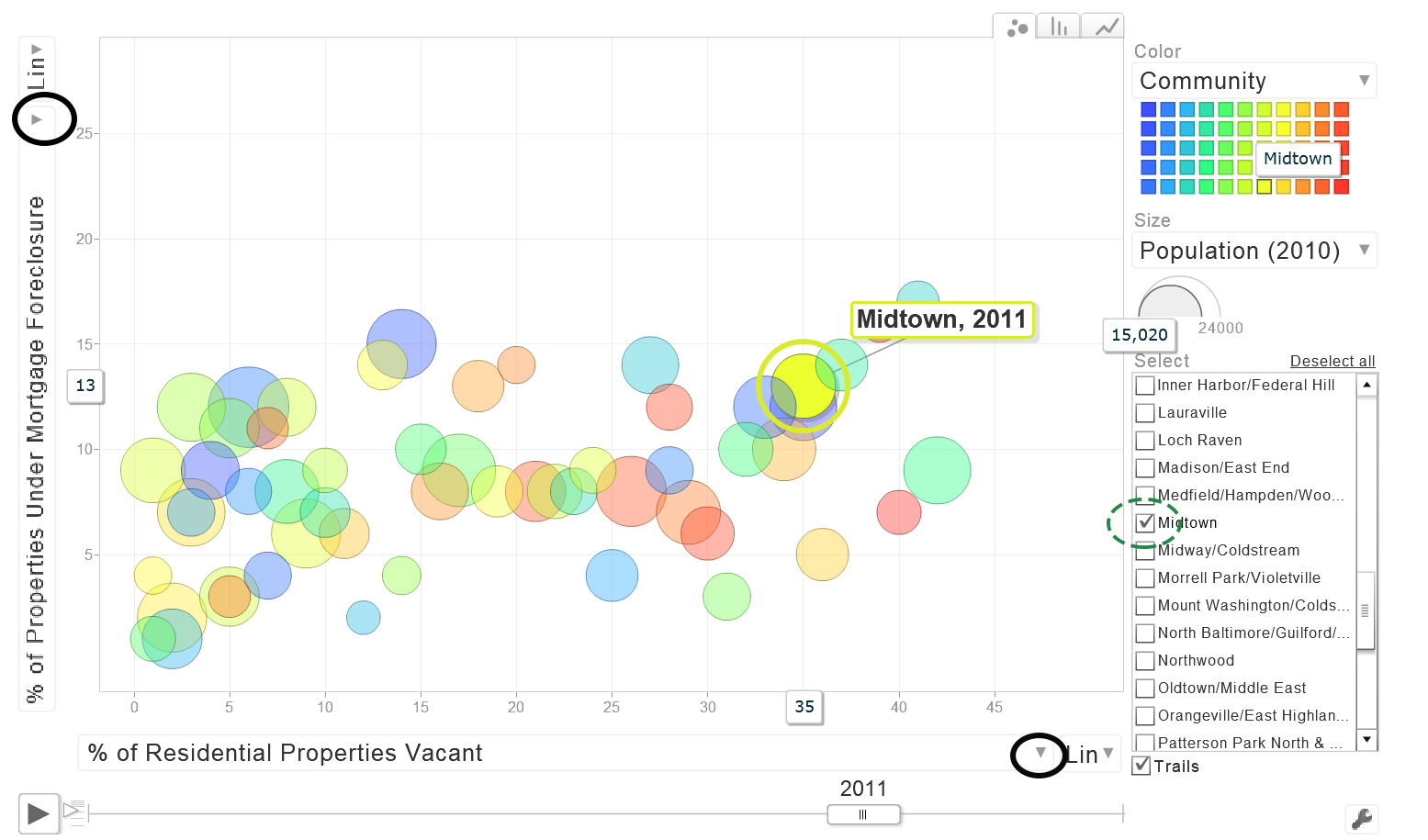}
	\end{center}
	\caption{Screen shot of the Baltimore housing variables motion chart, showing percent of residential properties vacant versus
		percent of properties under mortgage foreclosure. Variables to show can be selected from pull down menus that appear when the arrows are pressed (black circles). Selecting a community (green circle) and then running the animation will track the chosen community over time.} \label{fig:Baltimoremotionchartnewvar}
\end{figure}


\begin{figure}[H]
	\begin{center}

		\includegraphics[width=3.8in]{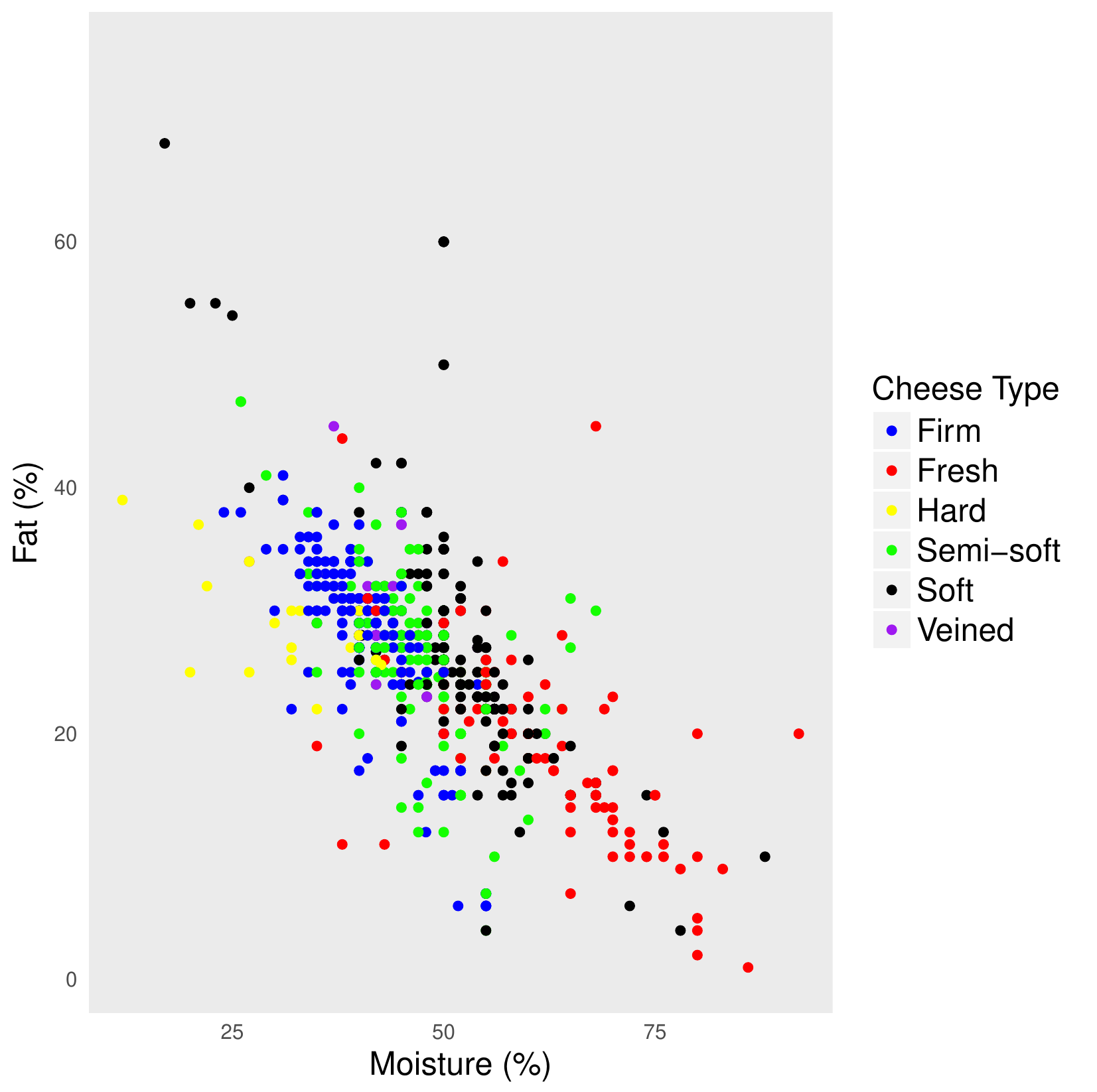}
	\end{center}
	\caption{Scatterplot of cheese fat percentage and moisture percentage color coded by cheese type, using Canada open data (https://open.canada.ca/en/open-data). It is hard to tell in this chart if cheese type has any influence in the association between fat percentage and moisture percentage.} \label{fig:spcheesecolor}
\end{figure}

\begin{figure}[H]
	\begin{center}

		\includegraphics[width=3.8in]{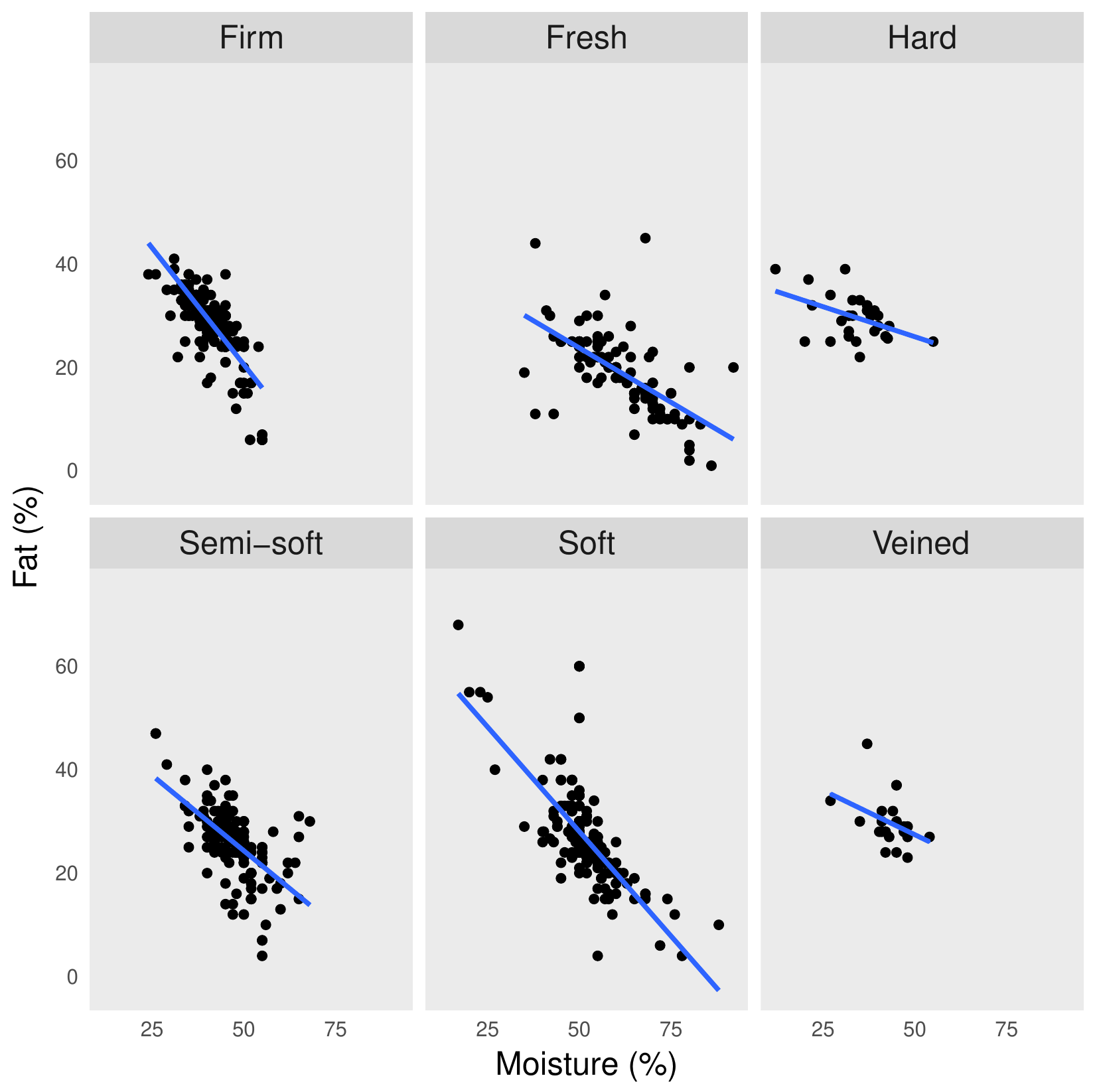}
	\end{center}
	\caption{Scatterplots of cheese fat percentage and moisture percentage per cheese type, using data from the Canada open data portal. It is easier to tell in this chart that cheese type has some influence in the association between fat percentage and moisture percentage. See supplementary material for reproducible \textsf{R} code.} \label{fig:spcheesepanel}
\end{figure}

Interactive plots have generated a lot of buzz. For one case study, a motion chart was created by combining several different data sets\footnote{See supplementary material.}. 
Users can select variables to compare and track how the measure of a location changes in time. While aesthetic appeal is important and certainly catches the attention of students during class, one of the key roles of visualization techniques in data analysis is to present information effectively. On occasion, attractive plots (Figure \ref{fig:spcheesecolor}) may not be as effective in providing information, as more basic visualization techniques of the same data (Figure \ref{fig:spcheesepanel}, where correlation coefficients, in alphabetic order of cheese type, are: -0.78, -0.64, -0.47, -0.62, -0.70, and -0.38).

The sample case studies presented in this article only scratch the surface of the possible ways open data can be incorporated into courses. The San Diego vehicle stops data presented in section \ref{sec:sdpol} could be used while discussing empirical probabilities or hypothesis testing on the dependence of two categorical variables. The Canadian cheese data could serve as a platform to teach multiple linear regression with interactions. Table \ref{tab:ideas} presents a few more ideas of statistical methods that can be taught using open data. The Chicago taxi ride data includes over 113 million records, and it is so large that it can not be opened with most statistical software (including \textsf{R}); requiring big data tools such as querying, subsetting, or the use of special 
software/hardware (\cite{baumeretal17}). The supplementary material provides code to subset the data. In advanced courses, a sophisticated model can be fit to the taxi ride data for the purpose of flexibility and accounting for all available predictors in the data set.   

\begin{table}[H]
	\footnotesize
	\centering
	\scalebox{0.7}{
	\begin{tabular}{cccc}
		\hline
		\textbf{Data} &  \textbf{Location} & \textbf{Topic} & \textbf{Comments} \\[5pt] 
		Lead in water                 &  Toronto              & Quartiles  &  Data is censored.   \\[5pt] 
		Emergency response time &  New York City & ANOVA & Large data makes it easier  \\
		for cardiac arrest  &&& to detect smaller differences among parameters. \\[5pt] 
		Legal marijuana sales &  Washington & Two population test & ANOVA on product type possible as well. \\[5pt]
		Fatal shootings & Philadelphia & Two proportion inference & Confidence interval or testing.\\[5pt]
		Taxi ride data & Chicago & Multiple linear regression & Over 113 million taxi rides.\\
		&&&                                                             Big data methods necessary.\\[5pt]
		Meteorite mass & World & Shape of distribution&  Choose specific meteorite classification.\\[5pt]
		Tourism employment & Canada & Time series & Seasonal component and global trend present.\\\hline            
	\end{tabular}
}
	\caption{A few more ideas for incorporating open data into introductory statistics. See supplementary material for reproducible \textsf{R} codes.}
	\label{tab:ideas}
\end{table}

Case Studies are not the only pedagogical tool that can be developed from open data. Class projects are also possible. Open data can also have a role in homeworks, quizzes or exams. For example, the class can be assigned a data set to perform a statistical procedure. Assignments can be individualized by asking pupils to take a sample of size $n$ based on, say the last two digits of the students identification number. For example, if a student's last two digits is 14 and $n=50$, then the student must perform the assignment using observations from row 14 to 63. This way, the instructor can create \textsf{R} code to efficiently reproduce each student sample and grade the assignments.

\section{Discussion}

The ubiquity of open data has the advantage of allowing instructors to easily `locally adapt' lesson plans; which would follow more faithfully the 2016 GAISE College Report recommendation to integrate real data with context and purpose. Specifically, there is crime data from Los Angeles, New York and other cities with which a case study analogous to section \ref{sec:prm} can be built. Similarly, vehicle stops data is available for Austin Texas and other locations. \textsf{R} code for the case studies presented can be found in the supplementary material. 

Another benefit of open data is that it is often raw and very large. Thus, students get a first look into data science and big data with such data sets. For example, the Chicago taxi ride raw data has over 113 million records; requiring big data methods (\cite{baumeretal17}). Case study 1 required combining several data sets, filtering, aggregation, and the creation of a new variable: murder rate (Figure \ref{fig:murderratesteps}). The supplementary material sheds light on these steps and the assessment of errors. Although we advocate performing data wrangling in class, we do not recommend its implementation on every single data set presented. Instead, we propose implementing basic data preprocessing a few times, and reminding students of its importance throughout the course. Understanding the role of data preprocessing is no less important for non-majors. It is unlikely that non-majors perform data science or statistical analysis themselves; but they could contribute in the creation of a cohesive data analysis process. The reader should refer to the supplementary \textsf{R} codes to see more examples of data wrangling. Although the emphasis in this article is introductory statistics courses, the applications proposed can be modified accordingly for more specialized courses.

There are some challenges to be aware of when incorporating open data into the classroom. Sometimes it is not clear where the data comes from, which puts the reliability of the data in question (\cite{rivera16}); or there is no variable dictionary available with the data set. Depending on the data set and context of a case study, parameters can be obtained, not
statistics (e.g. calculating October 2015 mean number of Orleans Parish Prison inmates, using their daily inmate data.). Some open data websites do not work properly on some web browsers. Furthermore, since a few data sets are routinely updated in open data portals, for certain tasks it is wise to ensure everyone is using the same version of the data by providing a downloaded version. Moreover, the data may hold hidden surprises. For example, lead measurements from sampled tap water in Toronto are available online. A potential application is regression by postal code, or ANOVA. However, the lead measurements are censored, requiring more advanced procedures for inference than seen in introductory statistics courses. Instructors are advised to carefully explore the data before using it in class. Part of the idea is that the data is realistically complicated, but not too complicated for academic use.

Making data available through open data portals offers valuable potential benefits. Use of open data sets for statistical analysis and to teach statistics courses can encourage open portal managers to share data in a more efficient way. In principle, enough academic interest in open data could lead to improvements in open portal platforms: more data, better accessibility, and faster updating of data.



\end{document}